\documentclass[prd,aps,twocolumn]{revtex4}
\usepackage[utf8]{inputenc}
\def\gr{$\gamma$-ray}
\usepackage{graphicx}
\usepackage{color}

\begin{document}

\title{Signatures of anisotropic diffusion around PeVatrons in 100 TeV gamma-ray data}
\author{G.~Giacinti$^{1,2,3}$, T.~Abounnasr$^{4}$, A.~Neronov$^{5,6}$  and D.~Semikoz$^{5}$}
\affiliation{$^1$Max-Planck-Institut f\"ur Kernphysik, Postfach 103980, 69029 Heidelberg, Germany}
\affiliation{$^2$Tsung-Dao Lee Institute, Shanghai Jiao Tong University, Shanghai 200240, P. R. China}
\affiliation{$^3$School of Physics and Astronomy, Shanghai Jiao Tong University, Shanghai 200240, P. R. China}
\affiliation{$^4$Institute for Cosmic Ray Research, The University of Tokyo, 5-1-5 Kashiwanoha, Kashiwa, Chiba 277-8582, Japan}
\affiliation{$^5$Université de Paris Cite, CNRS, Astroparticule et Cosmologie, F-75013 Paris, France}
\affiliation{$^6$ Laboratory of Astrophysics, Ecole Polytechnique Federale de Lausanne, CH-1015, Lausanne, Switzerland}
\begin{abstract}
    The Tibet AS$\gamma$ collaboration has reported a diffuse \gr\ emission signal from the Galactic Plane. We consider the possibility that the diffuse emission from the outer Galactic Plane at the highest energies is produced by cosmic rays spreading from a single supernova-type source either in the Local or Perseus arm of the Milky Way. We show that anisotropic diffusion of multi-PeV cosmic rays along the Galactic magnetic field can produce an extended source spanning ten(s) of degrees on the sky, with a flux-per-unit-solid-angle consistent with Tibet AS$\gamma$ measurements. Observations of this new type of very extended sources, and measurements of their morphologies, can be used to characterize the anisotropic diffusion of PeV cosmic rays in the Galactic magnetic field, and to constrain the locations and properties of past PeVatrons.
\end{abstract}
\maketitle

\section{Introduction}

The detections of diffuse \gr\ emission from the Galactic Plane by Tibet AS$\gamma$ experiment \cite{TibetDiffuse2021}, and of isolated \gr\ sources by HAWC \cite{2019arXiv190908609H} and LHAASO \cite{2021Natur.594...33C} have extended the energy frontier of astronomy into the previously unexplored Peta-electronvolt energy range. Charged particles, possibly protons and nuclei, producing PeV \gr s have energies in the $\sim 10$~PeV energy range, i.e. in the range of the "knee" of the cosmic ray spectrum \cite{Apel2013,Aartsen2019}. The two types of \gr\ signals: diffuse emission and fluxes from isolated sources, are certainly related to each other. Isolated sources inject multi-PeV cosmic rays with a yet-to-be-determined spectrum into the interstellar medium, while the propagation of these cosmic rays through the interstellar medium generates the diffuse \gr\ flux. 

The details of the relation between the properties of isolated sources of PeV cosmic rays, called "PeVatrons", and the properties of the cosmic ray spectrum in the interstellar medium are not clear. Isolated sources detected by HAWC and LHAASO typically have rather soft spectra $dN_\gamma/dE\propto E^{-\Gamma_\gamma}$ with slopes $\Gamma_\gamma\ge 3$ in the 0.1-1PeV energy range \cite{2019arXiv190908609H,2021Natur.594...33C}. The assumption that these spectra are produced by interactions of cosmic rays escaping from the sources into the interstellar medium suggests that the spectra of cosmic rays injected by these sources are softer than $E^{-3}$ in the PeV range. Diffusive escape of PeV cosmic rays from the Galactic cosmic ray halo is expected to further soften the slope of the cosmic ray spectrum. In this situation, it is difficult to match the observed cosmic ray spectrum slope $dN_{cr}/dE_{cr}\propto E_{cr}^{-\Gamma_{cr}}$ with $2.5<\Gamma_{cr}\sim 3$ with the LHAASO observation of a soft-spectrum source population. The solution to this puzzle may lie in the peculiarities of the escape of very-high-energy cosmic rays from their sources and/or of the propagation of multi-PeV cosmic rays in the Galactic magnetic field.

Understanding the processes of cosmic ray escape from their sources and of cosmic ray propagation through the interstellar medium may be facilitated if the \gr\ emission from "clouds" of escaped cosmic rays could be reliably identified around individual sources.  This is perhaps difficult in the GeV-to-TeV energy range where sources are numerous and diffusion of cosmic rays through the interstellar medium is slow. In this case, the extended \gr\ emission produced by cosmic rays spreading from individual sources may be difficult to trace on top of the much stronger collective diffuse emission produced by the entire source population \cite{Neronov:2012kz}. However, the number of sources capable of accelerating particles to much higher energies is lower, and cosmic ray diffusion is faster for higher energy particles. In these conditions, the diffuse \gr\ emission in the highest, $\sim$\,PeV, energy range may be dominated by the signal from cosmic rays spreading from just a few sources. In this case, the identification of the extended emission from cosmic rays spreading from an individual source and the study of the morphology of the diffuse emission signal may clarify the details of the processes of injection of cosmic rays from the source and of cosmic ray propagation through the interstellar medium. 

In the following, we explore if the hypothesis of superposition of diffuse emission from cosmic rays spreading from just a few single source(s) may explain the diffuse emission signal from the direction of outer Galactic Plane observed by Tibet AS$\gamma$. We model the evolution of the distribution of cosmic rays spreading from a single source through the ordered and turbulent Galactic magnetic fields, using a Monte-Carlo technique tracing individual trajectories of cosmic rays. We calculate the \gr\ signal from the decays of neutral pions produced by interactions of cosmic rays spreading around the source and compare the predicted morphology of the \gr\ signal with the Tibet AS$\gamma$ data.

\section{Possible source locations in the outer Milky Way disk}

\begin{figure}
    \includegraphics[width=\linewidth]{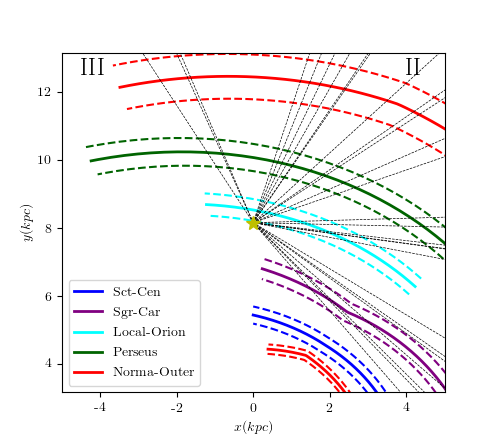}
    \caption{Milky Way spiral arm structure as defined in Ref. \cite{Reid_2019}, and zoomed-in around the Sun's position (yellow star). The dotted lines represent the directions (projected onto the Galactic plane) of the 398-1000\,TeV Tibet AS$\gamma$ events from~\cite{TibetDiffuse2021} with $|b| < 15^\circ$.}
    \label{fig:arms}
\end{figure}

To determine the most likely locations for individual PeVatrons in the outer Galaxy, we briefly review in this section the structure of the Milky Way disk and outline the model we use in the calculations of the subsequent sections. The Milky Way is a four-armed barred spiral galaxy \cite{2008gady.book.....B,2021A&A...651A.104P}. The Sun is located at the distance $D_\odot=8.15\pm0.15$~kpc \cite{Reid_2019} from the Galactic Center, in the immediate vicinity of the Orion --or "Local"-- arm, which lies between the Sun and the Perseus arm, see Fig.~\ref{fig:arms}.


An analytical model for the spiral arm structure can be defined as in Ref.~\cite{Reid_2019}:
\begin{equation}
    r(\theta)=R_{0}\exp{ \left[ \tan(p)\left(\theta - \theta_{0} \right) \right]}
\end{equation}
where ($r,\,\theta$) denote the polar coordinates in the Galactic disk, and $R_0$, $\theta_0$, and $p$ are parameters defined differently for the Perseus, Norma, Scrutum and Sagittarius arms. Following Ref.~\cite{Reid_2019}, we use two different values for the pitch angles $p$ of each arm. The arm structure and their scale widths can be found in table 2 of \cite{Reid_2019}. We plot this spiral arm structure in Fig.~\ref{fig:arms}, where the dashed lines correspond to the scale widths of the arms.

For the gas density in the interstellar medium, we adopt a simple model loosely based on the model presented in Ref.~\cite{Drimmel:2001ti}. In each spiral arm, we assume that the gas number density follows:
\begin{equation}
n_{ism}=n_{arm,0} \exp\left(-\frac{z^2}{2\sigma_z^2} \right)\,, \label{eq:2}
\end{equation}
where $n_{arm,0}$ is the mid-plane density for each arm, and $z$ is the distance to the Galactic plane. The vertical scale height of the arms $\sqrt{2}\sigma_z$ is chosen such that:
\begin{equation}
    \sigma_z=\left\{
    \begin{array}{ll}
    40 ~ \mathrm{pc}, &r< 7 ~ \mathrm{kpc}\\
    40 ~ \mathrm{pc} + 72 ~ \frac{\mathrm{pc}}{\mathrm{kpc}} ~ (r-7~\mathrm{kpc}),&r \geq 7 ~ \mathrm{kpc}\,,
    \end{array}
    \right. \label{eq:3}
\end{equation}
which is twice the rms value evaluated in~\cite{Reid_2019}.


The Galactic disk has a warp that can be described as a shift of the middle plane along the $z$ axis. We include this warp in our calculations, using the formula provided in Eq.~(1) of Ref.~\cite{Skowron_2019}. Viewed from the Sun's position, the presence of this warp leads to a slight shift of the galactic plane not larger than 2$^\circ$, and it only has a small effect in the region of the Galactic disk that we consider in Section 4.

\begin{figure}
    \includegraphics[width=\linewidth]{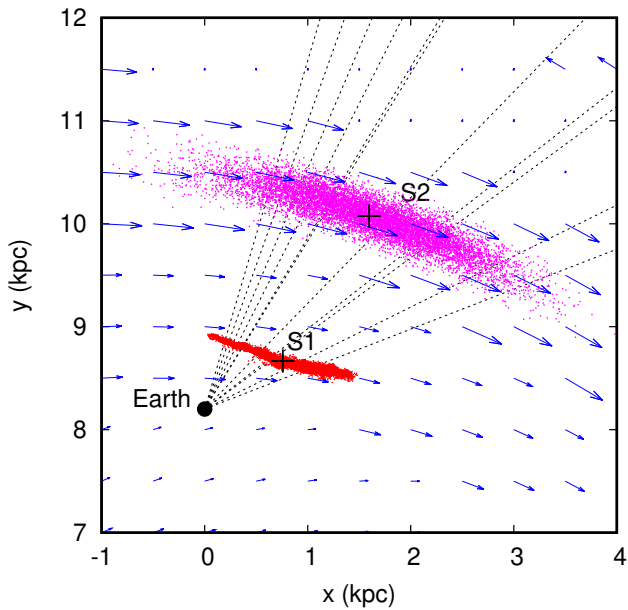}
    \caption{Sketch of a 5\,kpc$\times$5\,kpc region of interest of the Galactic plane, seen from above. The black dot represents the Earth's location, and the two black crosses are those of \lq\lq S1\rq\rq\/(Source in the local --Orion-- arm) and \lq\lq S2\rq\rq\/(Source in the Perseus arm). The red and magenta dots represent the locations, projected onto the Galactic plane, of the $10^4$ simulated cosmic rays that have escaped respectively from S1 and from S2.  The blue arrows show the directions of the regular Galactic magnetic field in the Jansson \& Farrar model~\cite{JF_GMF_1,JF_GMF_2}, and their sizes are proportional to the field strength. The thin black dotted lines represent the directions (projected onto the Galactic plane) of the 11 Tibet AS$\gamma$ events at $l \approx 110^\circ - 170^\circ$ and $|b| < 15^\circ$.}
    \label{fig:GalPlane}
\end{figure}

For the regular and turbulent components of the Galactic magnetic field, we adopt the Jansson \& Farrar model from Refs.~\cite{JF_GMF_1,JF_GMF_2}, with a rescaling of the strength of its turbulent component as explained in Section~4. In the Galactic disk, at $z=0$, the regular magnetic field is almost aligned with the spiral arm structure. In Fig.~\ref{fig:GalPlane}, we plot with blue arrows the direction and strength of this regular magnetic field at $z=0$. One can see that it follows globally the directions of the Local and Perseus arms from Fig.~\ref{fig:arms}.

In Fig.~\ref{fig:arms}, we show with dotted black lines the directions of the Tibet AS$\gamma$ events from~\cite{TibetDiffuse2021}, that are in the 398-1000\,TeV energy range and lie at Galactic latitudes $|b|<15^\circ$. These events likely correspond to the diffuse \gr\ emission from the Galactic plane. In Fig.~\ref{fig:GalPlane}, we only plot the directions of the 11 events that lie at $l = 110^\circ - 170^\circ$. One can clearly see that a multi-PeV cosmic ray source located in either the Local arm or the Perseus arm could in principle inject particles that would diffuse along the magnetic field in these arms and produce some of these 11 events. In the following two sections, we investigate analytically and numerically if such a scenario is viable.

\section{Propagation of cosmic rays from a source in the Local and Perseus arms and diffuse \gr\ emission}

Cosmic rays spiral along magnetic field lines and are scattered by the turbulent component of the Galactic magnetic field. 
The modelling of \cite{Giacinti:2017dgt} suggests that turbulent magnetic fields with a Kolmogorov spectrum extending up to the maximal scale $L_{max}\sim 100$~pc (with coherence length $L_{coh}=L_{max}/5\sim 25$~pc) assures diffusive propagation of cosmic rays in the energy range in which the gyroradius 
\begin{equation}
    R_L=\frac{E_p}{eB}\sim 0.3\left[\frac{E_p}{10^{15}\mbox{ eV}}\right]\left[\frac{B}{3\ \mu\mbox{G}}\right]^{-1}\mbox{ pc}
\end{equation}
is 
\begin{equation}
    R_L<\frac{L_{coh}}{2\pi}\simeq 4\left[\frac{L_{max}}{100\mbox{ pc}}\right]\mbox{ pc}
\end{equation}
Cosmic rays with energies up to 10~PeV (producing sub-PeV \gr\ emission in interactions with the interstellar medium) can propagate in the diffusive regime if the maximal scale  of the turbulence is $L_{max}\sim 10^2$~pc. 

Diffusion of particles in superimposed regular and turbulent magnetic fields is anisotropic with different diffusion coefficients $D_{||}, D_\bot$ parallel and perpendicular to the regular magnetic field direction \cite{Giacinti:2017dgt}. Both diffusion coefficients scale as a powerlaw in energy $D_{||}\propto E_p^{\delta_{||}}$, $D_\bot\propto E_p^{\delta_{\bot}}$ and numerical modelling indicates  that $\delta_{||}\simeq \delta_\bot$, so that we drop the indexes $||, \bot$ for $\delta$ \cite{Giacinti:2017dgt}. The assumption of turbulence with a Kolmogorov spectrum fixes $\delta$ to the value $\delta=1/3$. Numerically, we find that 
\begin{eqnarray}
    D_{\bot}&\simeq& 10^{28}\left[\frac{E_p}{1\mbox{ PeV}}\right]^\delta\mbox{cm}^2/\mbox{s}\nonumber\\
    D_{||}&\simeq& 10^{31}\left[\frac{E_p}{1\mbox{ PeV}}\right]^\delta\mbox{cm}^2/\mbox{s}
\end{eqnarray}
for our choice of parameters of magnetic field with the root-mean-square of the turbulent component and the regular field strength related as $B_{rms}/B_{reg}=0.5$.

Cosmic rays released from a source $T_s$ years ago spread to the distance 
\begin{eqnarray}
    d_{||}&\sim& \sqrt{D_{||}T_s}\simeq 1 \left[\frac{T_s}{10\mbox{ kyr}}\right]^{1/2}\left[\frac{E_p}{10\mbox{ PeV}}\right]^{1/6}\mbox{ kpc}\\
    d_\bot&\sim& \sqrt{D_\bot T_s}\simeq 30 \left[\frac{T_s}{10\mbox{ kyr}}\right]^{1/2}\left[\frac{E_p}{10\mbox{ PeV}}\right]^{1/6}\mbox{ pc}\nonumber
\end{eqnarray}
Cosmic rays contained in such ellipsoid-shaped regions (of dimensions $2d_{||}\times 2d_{\bot}\times 2d_\bot$) form an extended over-density on top of the average "sea" of cosmic rays from multiple sources accumulated on the time scale of escape from the Galactic disk. This time scale is long for cosmic rays with relatively low (GeV) energy, so that the overdensity created by a single source is difficult to spot. The shortening of the escape time scale with increasing cosmic ray energy leads to a lower level of the cosmic ray "sea" and a stronger overdensity due to a single source. 

It is possible to detect the ellipsoid-shaped overdensity of cosmic rays around a single source through the \gr\ and neutrino emission that it produces through interactions of cosmic rays with the interstellar medium. Multi-PeV cosmic rays produce \gr s with energies $E_\gamma\simeq 0.1 E_p$. As an example, the \gr\ emission from a region of projected size $2d_{||}\times 2d_\bot$ around a source situated at the distance $d_s\sim 2$~kpc in the Perseus arm is expected to span a region of angular size $\Delta_{||,\bot}=2d_{||,\bot}/d_s$,  
\begin{eqnarray}
\Delta_{||}&\simeq& 60^\circ\left[\frac{T_s}{10\mbox{ kyr}}\right]^{1/2}\left[\frac{E_\gamma}{0.1\mbox{ PeV}}\right]^{1/6}\left[\frac{d_s}{2\mbox{ kpc}}\right]^{-1}\nonumber\\
\Delta_\bot&\simeq& 2^\circ\left[\frac{T_s}{10\mbox{ kyr}}\right]^{1/2}\left[\frac{E_\gamma}{0.1\mbox{ PeV}}\right]^{1/6}\left[\frac{d_s}{2\mbox{ kpc}}\right]^{-1}
\end{eqnarray}
which can thus occupy a sizeable part of the Galactic Plane observed by Tibet AS$\gamma$. 

Cosmic rays loose energy in interactions with ambient gas on the time scale 
\begin{equation}
    t_{pp}=(\sigma_{pp}n_{ism})^{-1}\simeq 3\times 10^7\left[\frac{n_{ism}}{1\mbox{ cm}^{-3}}\right]^{-1}\mbox{ yr}
\end{equation}
where $\sigma_{pp}\simeq 3\times 10^{-26}\mbox{ cm}^2$ is the $pp$ interaction cross-section and $n_{ism}$ is the density of the interstellar medium.
These interactions produce the \gr\ luminosity
\begin{eqnarray} 
    &&L_\gamma=\frac{\kappa E_p(d{\cal E}_s/dE_p)}{t_{pp}}\simeq \\ && 10^{34}\left[\frac{\kappa}{0.1}\right]\left[\frac{E_p(d{\cal E}_s/dE_p)}{10^{49}\mbox{ erg}}\right]\left[\frac{n_{ism}}{1\mbox{ cm}^{-3}}\right]\mbox{ erg/s}\nonumber
\end{eqnarray}
where $E_p(d{\cal E}_e/dE_p)$ it the total energy of cosmic rays injected by the source in the (decade wide) PeV energy range  and $\kappa$ is the fraction of the cosmic ray energy transferred to \gr s in each interaction. 

The flux from a source at a distance $d_s$ is thus 
\begin{equation}
    F_\gamma\simeq 10^{-9}\left[\frac{\kappa}{0.1}\right]\left[\frac{E_p(d{\cal E}_s/dE_p)}{10^{49}\mbox{ erg}}\right]\left[\frac{n_{ism}}{1\mbox{ cm}^{-3}}\right]\left[\frac{d_s}{2\mbox{ kpc}}\right]^{-2}\frac{\mbox{GeV}}{\mbox{cm}^2\mbox{s}}
\end{equation}
The flux per unit solid angle is 
\begin{eqnarray}
    &&\frac{dF_\gamma}{d\Omega}\simeq   10^{-8}\left[\frac{\kappa}{0.1}\right]\left[\frac{E_p(d{\cal E}_s/dE_p)}{10^{49}\mbox{ erg}}\right]\left[\frac{n_{ism}}{1\mbox{ cm}^{-3}}\right]\nonumber\\ 
    &&\left[\frac{T_s}{10\mbox{ kyr}}\right]^{-1}\left[\frac{E_\gamma}{0.1\mbox{ PeV}}\right]^{-1/3}\frac{\mbox{GeV}}{\mbox{cm}^2\mbox{s sr}}
    \label{eq:per_sr}
\end{eqnarray}
One can notice that the flux per unit solid angle does not depend on the source distance. Further away sources produce a smaller overall flux, but span a smaller solid angle on the sky, so that the surface brightness remains the same. 

Comparing the estimate (\ref{eq:per_sr}) with the diffuse emission flux measured by Tibet AS$\gamma$ \cite{TibetDiffuse2021}, we find that the contribution from a single source, in which the overall energy injection in cosmic rays was comparable to that expected for a supernova, can explain the observed flux-per-solid angle in different parts of the Galactic Plane. The PeV energy range is unique in this respect: elongated cosmic ray "bubbles" produced by anisotropic diffusion of cosmic rays from individual sources are expected to produce a \gr\ flux comparable to the overall diffuse emission flux. At lower energies, the flux from individual elongated bubbles are the same as in the PeV band (assuming that individual sources inject cosmic rays with powerlaw spectra with $E^{-2}$ slope), but the overall level of diffuse emission (which is known to have a spectrum $\sim E^{-2.5}$ or softer \cite{2015arXiv150507601N,2017A&A...606A..22N,2020A&A...633A..94N}) is much higher than the single source flux.

\section{Numerical modelling }
\label{Modelling}

We confirm here the analytical estimates of the previous section with a numerical model based on the approach developed in Refs.~\cite{Giacintietal2014,Giacintietal2015}. It is a Monte-Carlo modelling based on the direct integration of the equations of motion of cosmic rays in Galactic magnetic field models. We demonstrate that at least 7 of the 11 events of Tibet AS$\gamma$ in the region ($l \approx 110^\circ - 170^\circ$, $|b| < 15^\circ$) may have been produced by one former PeVatron located in the Local or Perseus arm.

We use the Jansson \& Farrar model of \cite{JF_GMF_1,JF_GMF_2}, and the turbulent magnetic field is generated using the nested grid method from Ref.~\cite{Giacintietal2012}. The turbulence is of Kolmogorov type with $L_{\max} = 150$\,pc, i.e. $L_{\rm c} = 30$\,pc. Following our earlier findings in Refs.~\cite{Giacintietal2014,Giacintietal2015}, we reduce the strength of the turbulent field, $B_{\rm rms}$, by a factor $1/5$ compared to the values quoted in~\cite{JF_GMF_2}, so as to fit the Boron-to-Carbon ratio --The original values of $B_{\rm rms}$ from Ref.~\cite{JF_GMF_2} leading to a too large grammage and thereby to an overproduction of secondary nuclei.

We calculate the trajectories of $10^4$ cosmic rays injected in random directions from a point source and save the positions of these particles at fixed times after the injection instant. We use such fixed time snapshots to estimate the density of cosmic rays in a region around the source position. We then use this density estimate to calculate the \gr\ emission from cosmic ray interactions with the interstellar medium, using the AAFrag numerical code \cite{Kachelriess2019a,2021PhRvD.104l3027K}. The density of the target gas is given by Eqs.~(\ref{eq:2}) and~(\ref{eq:3}).

We consider two different source locations: a young nearby source located in the local Orion arm, which we denote as \lq\lq S1\rq\rq\/ in the following, and a more distant, older source located in the Perseus arm, which we denote as \lq\lq S2\rq\rq\/. The source S1 is located at $(x,y,z) = (0.758\,{\rm kpc},8.67\,{\rm kpc},0)$ in the coordinate system of Figs. \ref{fig:arms} and~\ref{fig:GalPlane}. Its age is $t=3$\,kyr. The source S2 is located at $(x,y,z) = (1.60\,{\rm kpc},10.1\,{\rm kpc},-250\,{\rm pc})$ and its age is $t=30$\,kyr. We need S2 to be located at such a distance below the Galactic plane to fit the Tibet AS$\gamma$ data. While this distance might seem large, it is reasonable and justified in view of the presence of high-mass stars at such distances from the Galactic plane in the Perseus arm: See the location of the black dots below the Galactic plane in the upper panel of Fig.~5 in Ref.~\cite{Reid_2019}.

Fig. \ref{fig:GalPlane} shows a scatter plot of locations of the cosmic rays injected by each source $3$~kyr and 30~kyr after the injection moment. One can see that, consistently with the estimates of the previous section, the regions occupied by the cosmic rays are ellipsoids elongated in the direction of the ordered magnetic field in the Local and Perseus arms, respectively.

\begin{figure*}[!t]
  \centerline{\includegraphics[width=0.75\textwidth]{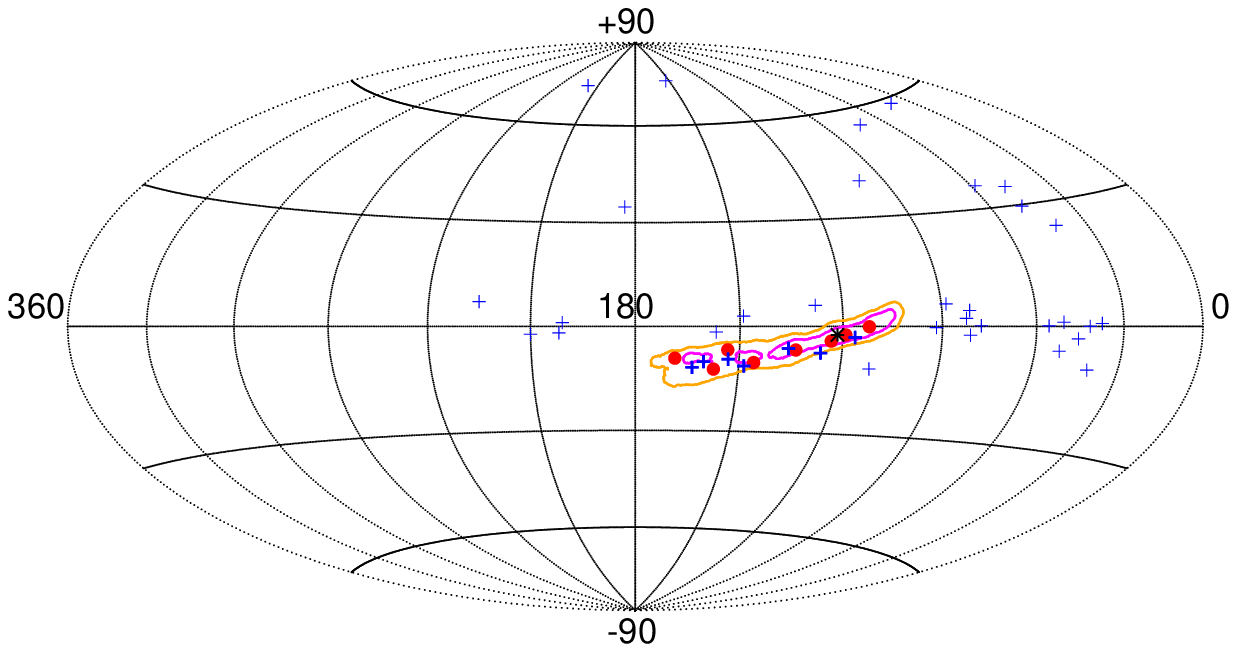}
             }
  \centerline{\includegraphics[width=0.75\textwidth]{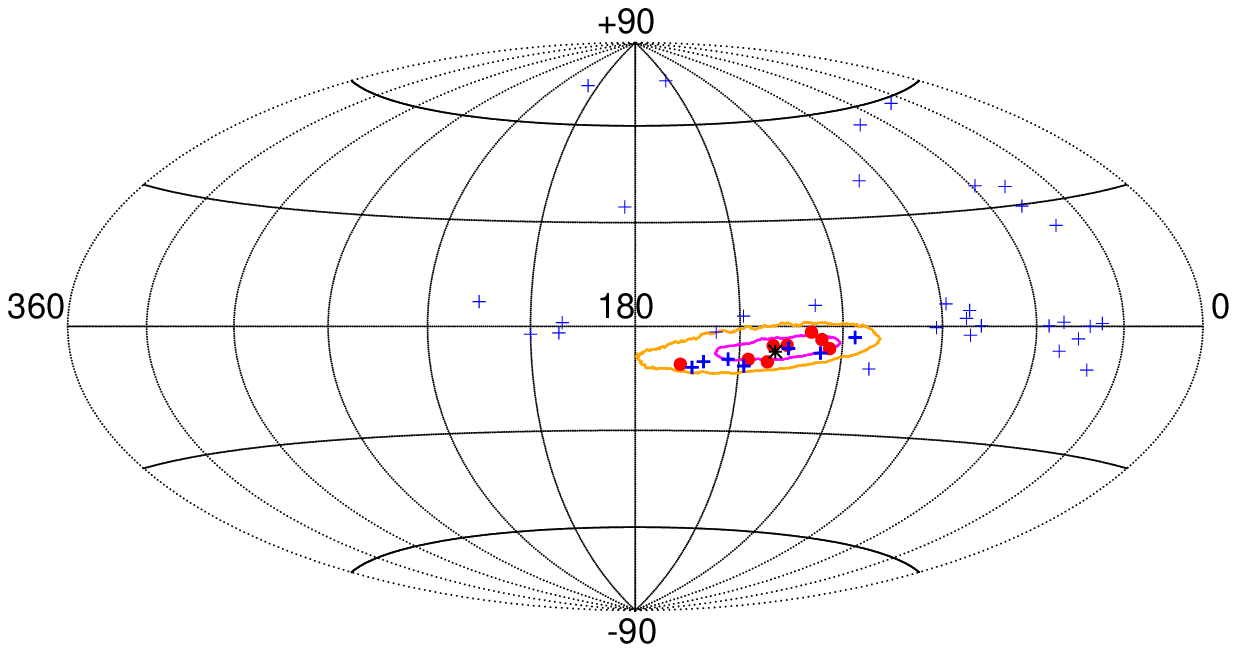}
             }
  \caption{Sky maps in Galactic coordinates showing the \gr\ emission in the 398--1000\,TeV band produced by the cosmic rays that escaped from S1 (upper panel) and S2 (lower panel). The black stars show the locations of S1 and S2 on the sky. The magenta and orange contours encircle the regions with the largest surface brightness, containing respectively 50\% and 95\% of the total \gr\ emission. The red dots show simulated random sets of gamma-ray events that would be detected in the 398--1000\,TeV energy range by a detector with Tibet AS$\gamma$'s exposure~\cite{TibetDiffuse2021}. The blue crosses represent the 38 Tibet AS$\gamma$ events in the 398--1000\,TeV energy range from Ref.~\cite{TibetDiffuse2021}. The 7 Tibet AS$\gamma$ events represented with thicker lines show the events falling within the source extensions.}
  \label{fig:Skymaps}
\end{figure*}

Fig. \ref{fig:Skymaps} shows the maps of the 398--1000\,TeV \gr\ emission resulting from the interaction of these cosmic rays with the gas in the interstellar medium (upper panel for S1, lower panel for S2). The contours correspond to the 50\% and 95\% signal containment. The arrival directions of the 38 Tibet AS$\gamma$ events with 398--1000\,TeV energies from Ref.~\cite{TibetDiffuse2021} are represented with blue crosses. There are 7 aligned events in the region $l \approx 110^\circ - 170^\circ$ and $-12^\circ<b<0$ (represented with thick blue crosses), which are all located within the 95\% signal containment contours, both for S1 and for S2. To compare our predicted \gr\ signal with the Tibet AS$\gamma$ data set, we simulate the \gr\ events statistics that would be observed with the exposure and observation time reported for this experiment in~\cite{TibetDiffuse2021}. Assuming that each source produces cosmic rays with an $E^{-2}$ spectrum and with a total energy $E_{\rm CR}$ between 1\,GeV and 10\,PeV, we reproduce the correct even statistics, namely $\simeq 7\pm 3$ photons in this energy range, for a mid-plane gas density $n_{arm,0}\simeq 0.33((10^{50}\,{\rm erg})/E_{\rm CR})$\,cm$^{-3}$ in the Local arm (S1), and $n_{arm,0}\simeq 1.5((3 \cdot 10^{50}\,{\rm erg})/E_{\rm CR})$\,cm$^{-3}$ in the Perseus arm (S2). The simulated events are shown with the red dots in Fig. \ref{fig:Skymaps}, for these values of target gas density. There are 8 events in both panels for the simulations shown here. The \gr\ event statistics predicted by our model is therefore consistent with that observed by Tibet AS$\gamma$, for the suggested source locations, and for standard mid-plane gas densities ($\sim 1$\,cm$^{-3}$) and typical supernova energies ($\sim 10^{51}$\,erg in kinetic energy, out of which $\approx 10$\% is channelled into cosmic rays). This demonstrates that these 7 aligned events in the data of Tibet AS$\gamma$ can be explained by one same cosmic ray source, located either in our local arm or in the Perseus arm, and with an energy compatible of that of a single supernova.

For other sources locations, between those of S1 and S2, the shape of the \gr\ emission may slightly change, but would still keep such an elongated shape. It is therefore possible that one or a few of the other 4 nearby Tibet AS$\gamma$ events at $l \approx 110^\circ - 170^\circ$ and $|b| < 15^\circ$ could be explained by the same source as the other 7 events.

\section{Discussion and conclusion}

Our calculations demonstrate that it is possible that the PeV-band diffuse emission from the Milky Way detected by Tibet AS$\gamma$ may come from just several individual very extended sources. Such sources are formed by cosmic rays spreading from the points of injection (for example, by individual recent supernovae) preferentially along the ordered, kpc-scale Galactic magnetic field. The existence of such a new class of very extended sources can be verified with higher-statistics observations of the PeV diffuse emission by the LHAASO Observatory \cite{2020PhRvD.102d3025N,Zhao:2021GJ} that will be able to map the diffuse \gr\ flux from both the Galactic disk and regions at higher Galactic latitudes in the near future. 

In the previous Section, we focused on the Tibet AS$\gamma$ events located in the region $l \approx 110^\circ - 170^\circ$ of the Galactic plane. We note that our model may also provide an explanation for the other two bundles of events in the $l \approx 60^\circ - 110^\circ$ region at $|b| < 15^\circ$. Two compact spots centered around $l \simeq 45^\circ$ and $l \simeq 85^\circ$ in the Galactic plane are clearly visible in Fig.~3. By comparing in Fig.~1 the arrival directions of these events (see the corresponding black dotted lines) to the Galactic spiral arm structure, one can see that the spots centered around $l \simeq 45^\circ$ and $l \simeq 85^\circ$ correspond, respectively, to the tangential directions to the Sagittarius arm and to the Local arm. Therefore, these two spots could be caused by one, or a few, extended sources in these arms. The fact that the regular Galactic magnetic field in these arms points approximately towards the Earth in these regions of the sky would explain why these spots appear more compact and less elongated than the source studied in Section~4: Their cosmic ray distributions would be more elongated along the line-of-sight, than across the sky.

Finding candidates for this new class of very extended sources in the data will also open a new way of exploring the geometry of the Galactic magnetic field. If a population of sources of this type is discovered, measurements of the directions along which these sources are elongated will trace the direction of the ordered Galactic magnetic field at different locations in the Galactic disk. 

Looking for very extended \gr\ sources of this type may also be the only possibility for discovering the elusive sources of the highest-energy ($\gtrsim$\,PeV) Galactic cosmic rays. The short escape time of these very-high-energy cosmic rays from their sources may preclude the possibility of identifying directly these sources through the interactions of the $\sim 10$\,PeV cosmic rays inside the sources, unless the acceleration event would be observed "in live", for example during the next Galactic supernova. The fast spread of $\sim 10$\,PeV cosmic rays through the interstellar medium also reduces the possibility of catching the signal from cosmic ray interactions in the immediate vicinity of the source (for example, in molecular clouds adjacent to the source). The detection of emissions extending on $\gtrsim 10^\circ$-scales and stretching along the ordered Galactic magnetic field may therefore provide a new and more robust way of identifying of the long-sought-after sources of 1-10\,PeV Galactic cosmic rays.

\bibliography{refs.bib}

\begin{thebibliography}{24}
\expandafter\ifx\csname natexlab\endcsname\relax\def\natexlab#1{#1}\fi
\expandafter\ifx\csname bibnamefont\endcsname\relax
  \def\bibnamefont#1{#1}\fi
\expandafter\ifx\csname bibfnamefont\endcsname\relax
  \def\bibfnamefont#1{#1}\fi
\expandafter\ifx\csname citenamefont\endcsname\relax
  \def\citenamefont#1{#1}\fi
\expandafter\ifx\csname url\endcsname\relax
  \def\url#1{\texttt{#1}}\fi
\expandafter\ifx\csname urlprefix\endcsname\relax\def\urlprefix{URL }\fi
\providecommand{\bibinfo}[2]{#2}
\providecommand{\eprint}[2][]{\url{#2}}

\bibitem[{\citenamefont{{Amenomori} et~al.}(2021)\citenamefont{{Amenomori},
  {Bao}, {Bi}, {Chen}, {Chen}, {Chen}, {Chen}, {Chen}, {Cirennima},
  {Danzengluobu} et~al.}}]{TibetDiffuse2021}
\bibinfo{author}{\bibfnamefont{M.}~\bibnamefont{{Amenomori}}},
  \bibinfo{author}{\bibfnamefont{Y.~W.} \bibnamefont{{Bao}}},
  \bibinfo{author}{\bibfnamefont{X.~J.} \bibnamefont{{Bi}}},
  \bibinfo{author}{\bibfnamefont{D.}~\bibnamefont{{Chen}}},
  \bibinfo{author}{\bibfnamefont{T.~L.} \bibnamefont{{Chen}}},
  \bibinfo{author}{\bibfnamefont{W.~Y.} \bibnamefont{{Chen}}},
  \bibinfo{author}{\bibfnamefont{X.}~\bibnamefont{{Chen}}},
  \bibinfo{author}{\bibfnamefont{Y.}~\bibnamefont{{Chen}}},
  \bibinfo{author}{\bibfnamefont{S.~W.} \bibnamefont{{Cirennima}},
  \bibfnamefont{Cui}}, \bibinfo{author}{\bibfnamefont{L.~K.}
  \bibnamefont{{Danzengluobu}}, \bibfnamefont{Ding}}, \bibnamefont{et~al.},
  \bibinfo{journal}{\prl} \textbf{\bibinfo{volume}{126}}, \bibinfo{eid}{141101}
  (\bibinfo{year}{2021}), \eprint{2104.05181}.

\bibitem[{\citenamefont{{HAWC Collaboration} et~al.}(2019)\citenamefont{{HAWC
  Collaboration}, {Abeysekara}, {Albert}, {Alfaro}, {Camacho},
  {Arteaga-Vel{\'a}zquez}, {Arunbabu}, {Avila Rojas}, {Ayala Solares},
  {Baghmanyan} et~al.}}]{2019arXiv190908609H}
\bibinfo{author}{\bibnamefont{{HAWC Collaboration}}},
  \bibinfo{author}{\bibfnamefont{A.~U.} \bibnamefont{{Abeysekara}}},
  \bibinfo{author}{\bibfnamefont{A.}~\bibnamefont{{Albert}}},
  \bibinfo{author}{\bibfnamefont{R.}~\bibnamefont{{Alfaro}}},
  \bibinfo{author}{\bibfnamefont{J.~R.~A.} \bibnamefont{{Camacho}}},
  \bibinfo{author}{\bibfnamefont{J.~C.} \bibnamefont{{Arteaga-Vel{\'a}zquez}}},
  \bibinfo{author}{\bibfnamefont{K.~P.} \bibnamefont{{Arunbabu}}},
  \bibinfo{author}{\bibfnamefont{D.}~\bibnamefont{{Avila Rojas}}},
  \bibinfo{author}{\bibfnamefont{H.~A.} \bibnamefont{{Ayala Solares}}},
  \bibinfo{author}{\bibfnamefont{V.}~\bibnamefont{{Baghmanyan}}},
  \bibnamefont{et~al.}, \bibinfo{journal}{arXiv e-prints}
  \bibinfo{eid}{arXiv:1909.08609} (\bibinfo{year}{2019}), \eprint{1909.08609}.

\bibitem[{\citenamefont{{Cao} et~al.}(2021)\citenamefont{{Cao}, {Aharonian},
  {An}, {Axikegu}, {Bai}, {Bao}, {Bastieri}, {Bi}, {Bi}, {Cai}
  et~al.}}]{2021Natur.594...33C}
\bibinfo{author}{\bibfnamefont{Z.}~\bibnamefont{{Cao}}},
  \bibinfo{author}{\bibfnamefont{F.~A.} \bibnamefont{{Aharonian}}},
  \bibinfo{author}{\bibfnamefont{Q.}~\bibnamefont{{An}}},
  \bibinfo{author}{\bibfnamefont{L.~X.} \bibnamefont{{Axikegu}},
  \bibfnamefont{Bai}}, \bibinfo{author}{\bibfnamefont{Y.~X.}
  \bibnamefont{{Bai}}}, \bibinfo{author}{\bibfnamefont{Y.~W.}
  \bibnamefont{{Bao}}},
  \bibinfo{author}{\bibfnamefont{D.}~\bibnamefont{{Bastieri}}},
  \bibinfo{author}{\bibfnamefont{X.~J.} \bibnamefont{{Bi}}},
  \bibinfo{author}{\bibfnamefont{Y.~J.} \bibnamefont{{Bi}}},
  \bibinfo{author}{\bibfnamefont{H.}~\bibnamefont{{Cai}}},
  \bibnamefont{et~al.}, \bibinfo{journal}{\nat} \textbf{\bibinfo{volume}{594}},
  \bibinfo{pages}{33} (\bibinfo{year}{2021}).

\bibitem[{\citenamefont{Apel et~al.}(2013)\citenamefont{Apel,
  Arteaga-Vel{\'{a}}zquez, Bekk, Bertaina, Bl{\"{u}}mer, Bozdog, Brancus,
  Cantoni, Chiavassa, Cossavella et~al.}}]{Apel2013}
\bibinfo{author}{\bibfnamefont{W.~D.} \bibnamefont{Apel}},
  \bibinfo{author}{\bibfnamefont{J.~C.} \bibnamefont{Arteaga-Vel{\'{a}}zquez}},
  \bibinfo{author}{\bibfnamefont{K.}~\bibnamefont{Bekk}},
  \bibinfo{author}{\bibfnamefont{M.}~\bibnamefont{Bertaina}},
  \bibinfo{author}{\bibfnamefont{J.}~\bibnamefont{Bl{\"{u}}mer}},
  \bibinfo{author}{\bibfnamefont{H.}~\bibnamefont{Bozdog}},
  \bibinfo{author}{\bibfnamefont{I.~M.} \bibnamefont{Brancus}},
  \bibinfo{author}{\bibfnamefont{E.}~\bibnamefont{Cantoni}},
  \bibinfo{author}{\bibfnamefont{A.}~\bibnamefont{Chiavassa}},
  \bibinfo{author}{\bibfnamefont{F.}~\bibnamefont{Cossavella}},
  \bibnamefont{et~al.}, \bibinfo{journal}{Astroparticle Physics}
  \textbf{\bibinfo{volume}{47}}, \bibinfo{pages}{54} (\bibinfo{year}{2013}),
  \eprint{1308.2098}.

\bibitem[{\citenamefont{Aartsen et~al.}(2019)\citenamefont{Aartsen, Ackermann,
  Adams, Aguilar, Ahlers, Ahrens, Alispach, Andeen, Anderson, Ansseau
  et~al.}}]{Aartsen2019}
\bibinfo{author}{\bibfnamefont{M.~G.} \bibnamefont{Aartsen}},
  \bibinfo{author}{\bibfnamefont{M.}~\bibnamefont{Ackermann}},
  \bibinfo{author}{\bibfnamefont{J.}~\bibnamefont{Adams}},
  \bibinfo{author}{\bibfnamefont{J.~A.} \bibnamefont{Aguilar}},
  \bibinfo{author}{\bibfnamefont{M.}~\bibnamefont{Ahlers}},
  \bibinfo{author}{\bibfnamefont{M.}~\bibnamefont{Ahrens}},
  \bibinfo{author}{\bibfnamefont{C.}~\bibnamefont{Alispach}},
  \bibinfo{author}{\bibfnamefont{K.}~\bibnamefont{Andeen}},
  \bibinfo{author}{\bibfnamefont{T.}~\bibnamefont{Anderson}},
  \bibinfo{author}{\bibfnamefont{I.}~\bibnamefont{Ansseau}},
  \bibnamefont{et~al.}, \bibinfo{journal}{Physical Review D}
  \textbf{\bibinfo{volume}{100}}, \bibinfo{pages}{82002}
  (\bibinfo{year}{2019}), \eprint{1906.04317}.

\bibitem[{\citenamefont{Neronov and Semikoz}(2012)}]{Neronov:2012kz}
\bibinfo{author}{\bibfnamefont{A.}~\bibnamefont{Neronov}} \bibnamefont{and}
  \bibinfo{author}{\bibfnamefont{D.~V.} \bibnamefont{Semikoz}},
  \bibinfo{journal}{Phys. Rev. D} \textbf{\bibinfo{volume}{85}},
  \bibinfo{pages}{083008} (\bibinfo{year}{2012}), \eprint{1201.1660}.

\bibitem[{\citenamefont{{Binney} and {Tremaine}}(2008)}]{2008gady.book.....B}
\bibinfo{author}{\bibfnamefont{J.}~\bibnamefont{{Binney}}} \bibnamefont{and}
  \bibinfo{author}{\bibfnamefont{S.}~\bibnamefont{{Tremaine}}},
  \emph{\bibinfo{title}{{Galactic Dynamics: Second Edition}}}
  (\bibinfo{year}{2008}).

\bibitem[{\citenamefont{{Poggio} et~al.}(2021)\citenamefont{{Poggio},
  {Drimmel}, {Cantat-Gaudin}, {Ramos}, {Ripepi}, {Zari}, {Andrae}, {Blomme},
  {Chemin}, {Clementini} et~al.}}]{2021A&A...651A.104P}
\bibinfo{author}{\bibfnamefont{E.}~\bibnamefont{{Poggio}}},
  \bibinfo{author}{\bibfnamefont{R.}~\bibnamefont{{Drimmel}}},
  \bibinfo{author}{\bibfnamefont{T.}~\bibnamefont{{Cantat-Gaudin}}},
  \bibinfo{author}{\bibfnamefont{P.}~\bibnamefont{{Ramos}}},
  \bibinfo{author}{\bibfnamefont{V.}~\bibnamefont{{Ripepi}}},
  \bibinfo{author}{\bibfnamefont{E.}~\bibnamefont{{Zari}}},
  \bibinfo{author}{\bibfnamefont{R.}~\bibnamefont{{Andrae}}},
  \bibinfo{author}{\bibfnamefont{R.}~\bibnamefont{{Blomme}}},
  \bibinfo{author}{\bibfnamefont{L.}~\bibnamefont{{Chemin}}},
  \bibinfo{author}{\bibfnamefont{G.}~\bibnamefont{{Clementini}}},
  \bibnamefont{et~al.}, \bibinfo{journal}{\aap} \textbf{\bibinfo{volume}{651}},
  \bibinfo{eid}{A104} (\bibinfo{year}{2021}), \eprint{2103.01970}.

\bibitem[{\citenamefont{Reid et~al.}(2019)\citenamefont{Reid, Menten,
  Brunthaler, Zheng, Dame, Xu, Li, Sakai, Wu, Immer et~al.}}]{Reid_2019}
\bibinfo{author}{\bibfnamefont{M.~J.} \bibnamefont{Reid}},
  \bibinfo{author}{\bibfnamefont{K.~M.} \bibnamefont{Menten}},
  \bibinfo{author}{\bibfnamefont{A.}~\bibnamefont{Brunthaler}},
  \bibinfo{author}{\bibfnamefont{X.~W.} \bibnamefont{Zheng}},
  \bibinfo{author}{\bibfnamefont{T.~M.} \bibnamefont{Dame}},
  \bibinfo{author}{\bibfnamefont{Y.}~\bibnamefont{Xu}},
  \bibinfo{author}{\bibfnamefont{J.}~\bibnamefont{Li}},
  \bibinfo{author}{\bibfnamefont{N.}~\bibnamefont{Sakai}},
  \bibinfo{author}{\bibfnamefont{Y.}~\bibnamefont{Wu}},
  \bibinfo{author}{\bibfnamefont{K.}~\bibnamefont{Immer}},
  \bibnamefont{et~al.}, \bibinfo{journal}{The Astrophysical Journal}
  \textbf{\bibinfo{volume}{885}}, \bibinfo{pages}{131} (\bibinfo{year}{2019}),
  ISSN \bibinfo{issn}{1538-4357},
  \urlprefix\url{http://dx.doi.org/10.3847/1538-4357/ab4a11}.

\bibitem[{\citenamefont{Drimmel and Spergel}(2001)}]{Drimmel:2001ti}
\bibinfo{author}{\bibfnamefont{R.}~\bibnamefont{Drimmel}} \bibnamefont{and}
  \bibinfo{author}{\bibfnamefont{D.~N.} \bibnamefont{Spergel}},
  \bibinfo{journal}{Astrophys. J.} \textbf{\bibinfo{volume}{556}},
  \bibinfo{pages}{181} (\bibinfo{year}{2001}), \eprint{astro-ph/0101259}.

\bibitem[{\citenamefont{Skowron et~al.}(2019)\citenamefont{Skowron, Skowron,
  Mróz, Udalski, Pietrukowicz, Soszyński, Szymański, Poleski, Kozlowski,
  Ulaczyk et~al.}}]{Skowron_2019}
\bibinfo{author}{\bibfnamefont{D.}~\bibnamefont{Skowron}},
  \bibinfo{author}{\bibfnamefont{J.}~\bibnamefont{Skowron}},
  \bibinfo{author}{\bibfnamefont{P.}~\bibnamefont{Mróz}},
  \bibinfo{author}{\bibfnamefont{A.}~\bibnamefont{Udalski}},
  \bibinfo{author}{\bibfnamefont{P.}~\bibnamefont{Pietrukowicz}},
  \bibinfo{author}{\bibfnamefont{I.}~\bibnamefont{Soszyński}},
  \bibinfo{author}{\bibfnamefont{M.}~\bibnamefont{Szymański}},
  \bibinfo{author}{\bibfnamefont{R.}~\bibnamefont{Poleski}},
  \bibinfo{author}{\bibfnamefont{S.}~\bibnamefont{Kozlowski}},
  \bibinfo{author}{\bibfnamefont{K.}~\bibnamefont{Ulaczyk}},
  \bibnamefont{et~al.}, \bibinfo{journal}{Acta Astronomica}
  \textbf{\bibinfo{volume}{69}} (\bibinfo{year}{2019}).

\bibitem[{\citenamefont{{Jansson} and {Farrar}}(2012{\natexlab{a}})}]{JF_GMF_1}
\bibinfo{author}{\bibfnamefont{R.}~\bibnamefont{{Jansson}}} \bibnamefont{and}
  \bibinfo{author}{\bibfnamefont{G.~R.} \bibnamefont{{Farrar}}},
  \bibinfo{journal}{\apj} \textbf{\bibinfo{volume}{757}}, \bibinfo{eid}{14}
  (\bibinfo{year}{2012}{\natexlab{a}}), \eprint{1204.3662}.

\bibitem[{\citenamefont{{Jansson} and {Farrar}}(2012{\natexlab{b}})}]{JF_GMF_2}
\bibinfo{author}{\bibfnamefont{R.}~\bibnamefont{{Jansson}}} \bibnamefont{and}
  \bibinfo{author}{\bibfnamefont{G.~R.} \bibnamefont{{Farrar}}},
  \bibinfo{journal}{\apjl} \textbf{\bibinfo{volume}{761}}, \bibinfo{eid}{L11}
  (\bibinfo{year}{2012}{\natexlab{b}}), \eprint{1210.7820}.

\bibitem[{\citenamefont{Giacinti et~al.}(2018)\citenamefont{Giacinti,
  Kachelriess, and Semikoz}}]{Giacinti:2017dgt}
\bibinfo{author}{\bibfnamefont{G.}~\bibnamefont{Giacinti}},
  \bibinfo{author}{\bibfnamefont{M.}~\bibnamefont{Kachelriess}},
  \bibnamefont{and} \bibinfo{author}{\bibfnamefont{D.~V.}
  \bibnamefont{Semikoz}}, \bibinfo{journal}{JCAP}
  \textbf{\bibinfo{volume}{07}}, \bibinfo{pages}{051} (\bibinfo{year}{2018}),
  \eprint{1710.08205}.

\bibitem[{\citenamefont{{Neronov} and {Malyshev}}(2015)}]{2015arXiv150507601N}
\bibinfo{author}{\bibfnamefont{A.}~\bibnamefont{{Neronov}}} \bibnamefont{and}
  \bibinfo{author}{\bibfnamefont{D.}~\bibnamefont{{Malyshev}}},
  \bibinfo{journal}{arXiv e-prints} \bibinfo{eid}{arXiv:1505.07601}
  (\bibinfo{year}{2015}), \eprint{1505.07601}.

\bibitem[{\citenamefont{{Neronov} et~al.}(2017)\citenamefont{{Neronov},
  {Malyshev}, and {Semikoz}}}]{2017A&A...606A..22N}
\bibinfo{author}{\bibfnamefont{A.}~\bibnamefont{{Neronov}}},
  \bibinfo{author}{\bibfnamefont{D.}~\bibnamefont{{Malyshev}}},
  \bibnamefont{and} \bibinfo{author}{\bibfnamefont{D.~V.}
  \bibnamefont{{Semikoz}}}, \bibinfo{journal}{\aap}
  \textbf{\bibinfo{volume}{606}}, \bibinfo{eid}{A22} (\bibinfo{year}{2017}),
  \eprint{1705.02200}.

\bibitem[{\citenamefont{{Neronov} and
  {Semikoz}}(2020{\natexlab{a}})}]{2020A&A...633A..94N}
\bibinfo{author}{\bibfnamefont{A.}~\bibnamefont{{Neronov}}} \bibnamefont{and}
  \bibinfo{author}{\bibfnamefont{D.}~\bibnamefont{{Semikoz}}},
  \bibinfo{journal}{\aap} \textbf{\bibinfo{volume}{633}}, \bibinfo{eid}{A94}
  (\bibinfo{year}{2020}{\natexlab{a}}), \eprint{1907.06061}.

\bibitem[{\citenamefont{{Giacinti} et~al.}(2014)\citenamefont{{Giacinti},
  {Kachelrie{\ss}}, and {Semikoz}}}]{Giacintietal2014}
\bibinfo{author}{\bibfnamefont{G.}~\bibnamefont{{Giacinti}}},
  \bibinfo{author}{\bibfnamefont{M.}~\bibnamefont{{Kachelrie{\ss}}}},
  \bibnamefont{and} \bibinfo{author}{\bibfnamefont{D.~V.}
  \bibnamefont{{Semikoz}}}, \bibinfo{journal}{\prd}
  \textbf{\bibinfo{volume}{90}}, \bibinfo{eid}{041302} (\bibinfo{year}{2014}),
  \eprint{1403.3380}.

\bibitem[{\citenamefont{{Giacinti} et~al.}(2015)\citenamefont{{Giacinti},
  {Kachelrie{\ss}}, and {Semikoz}}}]{Giacintietal2015}
\bibinfo{author}{\bibfnamefont{G.}~\bibnamefont{{Giacinti}}},
  \bibinfo{author}{\bibfnamefont{M.}~\bibnamefont{{Kachelrie{\ss}}}},
  \bibnamefont{and} \bibinfo{author}{\bibfnamefont{D.~V.}
  \bibnamefont{{Semikoz}}}, \bibinfo{journal}{\prd}
  \textbf{\bibinfo{volume}{91}}, \bibinfo{eid}{083009} (\bibinfo{year}{2015}),
  \eprint{1502.01608}.

\bibitem[{\citenamefont{{Giacinti} et~al.}(2012)\citenamefont{{Giacinti},
  {Kachelrie{\ss}}, {Semikoz}, and {Sigl}}}]{Giacintietal2012}
\bibinfo{author}{\bibfnamefont{G.}~\bibnamefont{{Giacinti}}},
  \bibinfo{author}{\bibfnamefont{M.}~\bibnamefont{{Kachelrie{\ss}}}},
  \bibinfo{author}{\bibfnamefont{D.~V.} \bibnamefont{{Semikoz}}},
  \bibnamefont{and} \bibinfo{author}{\bibfnamefont{G.}~\bibnamefont{{Sigl}}},
  \bibinfo{journal}{\jcap} \textbf{\bibinfo{volume}{2012}}, \bibinfo{eid}{031}
  (\bibinfo{year}{2012}), \eprint{1112.5599}.

\bibitem[{\citenamefont{Kachelrie{\ss}
  et~al.}(2019)\citenamefont{Kachelrie{\ss}, Moskalenko, and
  Ostapchenko}}]{Kachelriess2019a}
\bibinfo{author}{\bibfnamefont{M.}~\bibnamefont{Kachelrie{\ss}}},
  \bibinfo{author}{\bibfnamefont{I.~V.} \bibnamefont{Moskalenko}},
  \bibnamefont{and}
  \bibinfo{author}{\bibfnamefont{S.}~\bibnamefont{Ostapchenko}},
  \bibinfo{journal}{Computer Physics Communications}
  \textbf{\bibinfo{volume}{245}}, \bibinfo{pages}{106846}
  (\bibinfo{year}{2019}), \eprint{1904.05129}.

\bibitem[{\citenamefont{{Koldobskiy} et~al.}(2021)\citenamefont{{Koldobskiy},
  {Kachelrie{\ss}}, {Lskavyan}, {Neronov}, {Ostapchenko}, and
  {Semikoz}}}]{2021PhRvD.104l3027K}
\bibinfo{author}{\bibfnamefont{S.}~\bibnamefont{{Koldobskiy}}},
  \bibinfo{author}{\bibfnamefont{M.}~\bibnamefont{{Kachelrie{\ss}}}},
  \bibinfo{author}{\bibfnamefont{A.}~\bibnamefont{{Lskavyan}}},
  \bibinfo{author}{\bibfnamefont{A.}~\bibnamefont{{Neronov}}},
  \bibinfo{author}{\bibfnamefont{S.}~\bibnamefont{{Ostapchenko}}},
  \bibnamefont{and} \bibinfo{author}{\bibfnamefont{D.~V.}
  \bibnamefont{{Semikoz}}}, \bibinfo{journal}{\prd}
  \textbf{\bibinfo{volume}{104}}, \bibinfo{eid}{123027} (\bibinfo{year}{2021}),
  \eprint{2110.00496}.

\bibitem[{\citenamefont{{Neronov} and
  {Semikoz}}(2020{\natexlab{b}})}]{2020PhRvD.102d3025N}
\bibinfo{author}{\bibfnamefont{A.}~\bibnamefont{{Neronov}}} \bibnamefont{and}
  \bibinfo{author}{\bibfnamefont{D.}~\bibnamefont{{Semikoz}}},
  \bibinfo{journal}{\prd} \textbf{\bibinfo{volume}{102}}, \bibinfo{eid}{043025}
  (\bibinfo{year}{2020}{\natexlab{b}}), \eprint{2001.11881}.

\bibitem[{\citenamefont{Zhao et~al.}(2021)\citenamefont{Zhao, Zhang, Zhang, and
  Yuan}}]{Zhao:2021GJ}
\bibinfo{author}{\bibfnamefont{S.}~\bibnamefont{Zhao}},
  \bibinfo{author}{\bibfnamefont{R.}~\bibnamefont{Zhang}},
  \bibinfo{author}{\bibfnamefont{Y.}~\bibnamefont{Zhang}}, \bibnamefont{and}
  \bibinfo{author}{\bibfnamefont{Q.}~\bibnamefont{Yuan}},
  \bibinfo{journal}{PoS} \textbf{\bibinfo{volume}{ICRC2021}},
  \bibinfo{pages}{859} (\bibinfo{year}{2021}).

\end{thebibliography}

\end{document}